\documentclass[a4paper,11pt]{article}
\usepackage{pos}
\usepackage{amsmath}

\usepackage{ytableau}
\ytableausetup{smalltableaux}
\RequirePackage[vcentermath]{youngtab}

\newcommand{\be}{\begin{equation}}
    \newcommand{\ee}{\end{equation}}
\newcommand{\bea}{\begin{eqnarray}}
    \newcommand{\eea}{\end{eqnarray}}

\title{$\mathcal{N}=2$ higher-spin theories and harmonic superspace}

\author[a]{Ioseph Buchbinder}
\author[a,b]{Evgeny~Ivanov}
\author*[a,b]{Nikita Zaigraev}

\affiliation[a]{Bogoliubov Laboratory of Theoretical Physics, JINR,\\
  141980 Dubna, Moscow region, Russia}

\affiliation[b]{Moscow Institute of Physics and Technology,\\
141700 Dolgoprudny, Moscow region, Russia}

\emailAdd{buchbinder@theor.jinr.ru}
\emailAdd{eivanov@theor.jinr.ru}
\emailAdd{nikita.zaigraev@phystech.edu}

\abstract{A brief review of the harmonic superspace approach to the construction of  $\mathcal{N}=2$ supersymmetric higher spin theories is given.
We define off-shell analytic harmonic gauge potentials of $\mathcal{N}=2$ supersymmetric higher-spin multiplets and of $\mathcal{N}=2$ superconformal higher-spin multiplets
for an arbitrary integer highest spin. The component contents of these $\mathcal{N}=2$ higher-spin supermultiplets
        are explicitly presented.
We also construct their cubic couplings to the hypermultiplet.
This short report summarizes the basic results of the series of published papers \cite{Buchbinder:2021ite, Buchbinder:2022kzl, Buchbinder:2022vra}
as well as announces those of the forthcoming article \cite{2024}.}

\FullConference{International Conference  on  Particle Physics and Cosmology 
    (professor V.A. Rubakov memorial conference, CPPCRubakov2023)
    \\
 02-07, October 2023\\
Yerevan, Armenia\\}

 \tableofcontents
\begin{document}
	 \renewcommand{\hookAfterAbstract}{%
		\par\bigskip
		\textsc{ArXiv ePrint}:
		\href{https://arxiv.org/abs/2402.05704}{2402.05704}
	}
\maketitle

\section{Introduction}

Construction of consistent interacting higher-spin theories is a long-standing problem. Various no-go theorems \cite{Bekaert:2010hw} indicate that setting up
self-consistent couplings of Fronsdal fields on a flat background  is a non-trivial task which requires a significant revision of the principles
and foundations of relativistic field theory. On the other hand, there are well-known nonlinear Vasiliev equations on an (A)dS background \cite{Vasiliev:1990en}
and the conformal theory of interacting higher spins on a flat background \cite{Segal:2002gd}. Of great interest are supersymmetric generalizations of higher spins.
Combining supersymmetry with gauge symmetry of higher spins provides significant limitations
on the structure of admissible interactions. As is well known, it is a superfield language that essentially simplifies the construction of supersymmetric theories.

In four dimensions, off-shell superfield approaches are known for theories with $\mathcal{N}\leq 3$. Most appropriate for construction of unconstrained off-shell formulations
of $\mathcal{N}=2$ supersymmetric theories is $\mathcal{N}=2$ harmonic superspace (HSS) \cite{18}. In a series of articles \cite{Buchbinder:2021ite, Buchbinder:2022kzl, Buchbinder:2022vra, 2024},
we showed that the HSS approach is very efficient for describing $\mathcal{N}=2$ massless Fronsdal fields \cite{Buchbinder:2021ite}, their cubic interactions
with a hypermultiplet \cite{Buchbinder:2022kzl, Buchbinder:2022vra} and higher-spin $\mathcal{N}=2$ superconformal multiplets \cite{2024}.
The purpose of this short article is to summarize the results obtained, so as to help the interested reader to get
familiar with these new applications of the HSS method. We also announce a number of results on $\mathcal{N}=2$ superconformal higher spins to be soon published.


\section{ABC of harmonic superspace}
\label{sec: 2}

The conventional $\mathcal{N}=2, 4D$ superspace is parametrized by the coordinates $\left(x^{\alpha\dot{\alpha}}, \theta^{\alpha i}, \bar{\theta}^{\dot{\alpha}i}\right)$, $i=1,2$.
In $\mathcal{N}=2$ harmonic superspace these coordinates are supplemented by new auxiliary harmonic variables $u^\pm_i$, which parametrize an internal  sphere $S^2$ and satisfy the condition $u^{+i}u^-_i = 1$.
Because of the presence of the $S^2$ harmonics, any harmonic superfield has an infinite number of component fields. One can consider the analytic basis in HSS,
$  \theta^{\pm \alpha, \dot\alpha} := \theta^{\alpha, \dot\alpha i} u^\pm_i,$ $x^{\alpha\dot{\alpha}}_A := x^{\alpha\dot{\alpha}} - 4i\theta^{\alpha(i} \bar\theta^{\dot{\alpha}j)}u^+_iu^-_j$.
$\mathcal{N}=2$ supersymmetry with parameters $\epsilon^{\alpha i}, \bar{\epsilon}^{\dot{\alpha}i}$
($\epsilon^{\pm\alpha} = \epsilon^{\alpha i}u^\pm_i$, $\bar{\epsilon}^{\pm\dot{\alpha}} = \bar{\epsilon}^{\dot{\alpha} i}u^\pm_i$) is realized on the HSS coordinates as:
\begin{equation}\label{eq: susy1}
    \delta_\epsilon x_A^{\alpha\dot{\alpha}} = -4i \left(\epsilon^{-\alpha}\bar{\theta}^{+\dot{\alpha}} + \theta^{+\alpha}\bar{\epsilon}^{-\dot{\alpha}} \right),
    \quad
    \delta_\epsilon \theta^{\pm \mu} = \epsilon^{\pm \mu},
    \quad
    \delta_\epsilon \bar{\theta}^{\pm \dot{\mu}} = \bar{\epsilon}^{\pm \dot{\mu}},
    \qquad
    \delta_\epsilon u^{\pm}_i = 0.
\end{equation}
It immediately follows from the form of these transformations that there is an invariant subspace with half of Grassmann variables:
\begin{equation}
    \zeta := \left( x_A^{\alpha\dot{\alpha}}, \theta^{+\alpha}, \bar{\theta}^{+\dot{\alpha}}, u^{\pm}_i \right).
\end{equation}
This is called \textit{analytic harmonic superspace}, and in  $\mathcal{N}=2$ theories it plays  a fundamental role similar to $\mathcal{N}=1$ chiral superspace in $\mathcal{N}=1$ theories.
All $\mathcal{N}=2$ multiplets admit a nice description by analytic superfields.
For description of the matter $\mathcal{N}=2$ hypermultiplets with non-zero mass (see below) it is also necesary to add an auxiliary coordinate $x^5$, such that:
\begin{equation}\label{eq: susy2}
    \delta_\epsilon x^5 = 2i \left(\epsilon^{-\alpha}\theta^+_\alpha - \bar{\epsilon}^-_{\dot{\alpha}} \bar{\theta}^{+\dot{\alpha}} \right).
\end{equation}
The analogue of complex conjugation in HSS is the \textit{tilde conjugation}. On complex functions, it acts like ordinary complex conjugation. The action of tilde-conjugation on the harmonic superspace
coordinates  is given by:
\begin{equation}
    \widetilde{x^{\alpha\dot{\alpha}}_A} = x^{\alpha\dot{\alpha}}_A,
    \quad
    \widetilde{\theta^\pm_\alpha} = \bar{\theta}^\pm_{\dot{\alpha}},
    \quad
    \widetilde{\bar{\theta}^\pm_{\dot{\alpha}}} = - \theta^\pm_\alpha,
    \quad
    \widetilde{u^{\pm i}} = - u_i^\pm,
    \quad
    \widetilde{u^\pm_i} = u^{\pm i},
    \quad
    \widetilde{x^5} = x^5.
\end{equation}

The important ingredients of the harmonic superspace formalism are the covariant harmonic derivatives. They are defined as follows:
\begin{subequations}
    \begin{equation}
        \mathcal{D}^{++} := \partial^{++} - 4i \theta^{+\rho} \bar{\theta}^{+\dot{\rho}} \partial_{\rho\dot{\rho}} + \theta^{+\hat{\rho}} \partial^+_{\hat{\rho}}
        + \left[(\theta^{+})^2 - (\bar{\theta}^{+})^2 \right] \partial_5,
    \end{equation}
    \begin{equation}
        \mathcal{D}^{--} := \partial^{--} - 4i \theta^{-\rho} \bar{\theta}^{-\dot{\rho}} \partial_{\rho\dot{\rho}} + \theta^{-\hat{\rho}} \partial^-_{\hat{\rho}}
        + \left[(\theta^{-})^2 - (\bar{\theta}^{-})^2 \right] \partial_5,
    \end{equation}
    \begin{equation}\label{eq:D0}
        \mathcal{D}^0 = \partial^0 + \theta^{+\hat{\rho}} \partial^-_{\hat{\rho}}
        -
        \theta^{-\hat{\rho}} \partial^+_{\hat{\rho}}
    \end{equation}
\end{subequations}
and satisfy $SU(2)$ algebraic relations:
\begin{equation}
    [\mathcal{D}^{++}, \mathcal{D}^{--}] = \mathcal{D}^0,
    \qquad
    [\mathcal{D}^0, \mathcal{D}^{\pm\pm}] = \pm 2 \mathcal{D}^{\pm\pm}.
\end{equation}
Here we used the standard notation for the partial harmonic derivatives:
\begin{equation}
    \partial^{++} = u^{+i} \frac{\partial}{\partial u^{-i}},
    \qquad
    \partial^{--} = u^{-i} \frac{\partial}{\partial u^{+i}},
    \qquad
    \partial^0 =  u^{+i} \frac{\partial}{\partial u^{+i}}
    -
    u^{-i} \frac{\partial}{\partial u^{-i}}.
\end{equation}

All known $\mathcal{N}=2$ theories have a superfield formulation in the harmonic superspace \cite{18}. Here we present
two important examples: the ultimate $\mathcal{N}=2$ multiplet of matter - a \textit{hypermultiplet} and the simplest \textit{Maxwell gauge multiplet}.

\textbf{\textit{Hypermultiplet}} is described by an unconstrained analytic superfield $q^+(\zeta)$. It contains a doublet of complex scalars $f^i$ and a pair of singlet spinors $\psi_\alpha, \kappa_\alpha$,
as well as an infinite set of auxiliary fields which comes from the harmonic $S^2$ expansions:
\begin{equation}
    q^{+}(\zeta) = f^iu^+_i + \theta^{+\alpha}\psi_\alpha + \bar{\theta}^+_{\dot{\alpha}} \bar{\kappa}^{\dot{\alpha}} + \text{auxiliary fields}.
\end{equation}
So the off-shell hypermultiplet carries $\mathbf{\boldsymbol{\infty}_B+ \boldsymbol{\infty}_F}$ degrees of freedom.
For massive hypermultiplet (with mass equal to central charge) one must introduce dependence on $x^5$ as $q^+(\zeta,x_5) = e^{imx^5} q^{+}(\zeta)$.
The free hypermultiplet action has the form\footnote{Integration measure in the analytic superspace is defined as $d\zeta^{(-4)}:= d^4x d^2\theta^+ d^2\bar{\theta}^+du$.
Harmonic integration is defined as $\int du\, 1 = 1$, otherwise $0$.}:
\begin{equation}\label{eq: hyper free}
    S_{free} = -\int d\zeta^{(-4)} \tilde{q}^+ \mathcal{D}^{++} q^+ = - \frac{1}{2} \int d\zeta^{(-4)} \, q^{+a} \mathcal{D}^{++} q^+_a.
\end{equation}
Here we introduced the notation $q^{+a} = (\tilde{q}^+, q^+)$, $q^+_a = \epsilon_{ab} q^{+b} = (q^+, -\tilde{q}^+)$. In the second form we have
manifest Pauli-G\"{u}rsey $SU(2)_{PG}$ symmetry (see e.g. \cite{18}). This form is most convenient for construction of higher-spin cubic vertices. After
eliminating an infinite number of auxiliary fields, the action \eqref{eq: hyper free} is reduced to the sum of free actions for a doublet of scalars and a pair of fermions,
so on shell there survives $4_B+4_F$ degrees of freedom. Note that both massive and massless hypermultiplets are described uniformly.

\textbf{\textit{$\mathcal{N}=2$ Maxwell supermultiplet}} is described by an unconstrained analytic gauge potential $V^{++}(\zeta)$ with the gauge freedom
$\delta_\lambda V^{++}(\zeta) = \mathcal{D}^{++} \lambda(\zeta)$. Using this freedom one can impose Wess-Zumino gauge:
{\small \begin{equation}
    V^{++}_{WZ} = -4i \theta^+_\beta \bar{\theta}^+_{\dot{\beta}} A^{\beta\dot{\beta}}
    + i(\bar{\theta}^+)^2 \phi - i (\theta^+)^2 \bar{\phi}
    + (\bar{\theta}^+)^2 \theta^{+\beta} \psi_\beta^i u^-_i + (\theta^+)^2 \bar{\theta}^{+\dot{\beta}}\bar{\psi}_{\dot{\beta}}^i u^-_i
    +
    (\theta^+)^2 (\bar{\theta}^+)^2 D^{(ij)}u^-_i u^-_j.
\end{equation}  }
Here we are left with the gauge spin $1$ field  $A_{\beta\dot{\beta}}$ with the gauge freedom $\delta A_{\beta\dot{\beta}}\sim \partial_{\beta\dot{\beta}}a$,
the doublet of fermions $\psi^i_\beta$, a complex scalar $\phi$ and an auxiliary triplet of real scalar fields $D^{(ij)}$. So off shell there are $\mathbf{8_B+8_F}$ degrees of freedom.
The free supersymmetric and gauge-invariant action reads:
\begin{equation}\label{eq: spin 1 action}
    S_{(s=1)}= \int d^4x d^8\theta du \, V^{++} V^{--},
\end{equation}
where $V^{--}$ is a solution of zero-curvature condition:
\begin{equation}
    \mathcal{D}^{++} V^{--} = \mathcal{D}^{--} V^{++}.
\end{equation}
In Wess-Zumino gauge, after elimination of auxiliary fields, the action is reduced to the sum of free Maxwell action, free action for doublet of fermions and
Klein-Gordon action for complex scalar.

\section{$\mathcal{N}=2$ higher-spin multiplets}
\label{sec: 3}

$\mathcal{N}=2$ higher-spin supermultiplet with highest spin $s\geq2$ (we will denote such multiplet as spin $\mathbf{s}$ supermultiplet) is described by the set of unconstrained  analytic gauge potentials
\footnote{Spin $\mathbf{1}$ also fits in this construction: we just have to omit all the superfields where the number of indices is formally negative.}:
 \begin{equation}\label{eq: non-conf prepot}
h^{++\alpha(s-1)\dot\alpha(s-1)}(\zeta),\;\;\; h^{++\alpha(s-2)\dot\alpha(s-2)}(\zeta),\;\;\; h^{++\alpha(s-1)\dot\alpha(s-2)+}(\zeta),\;\;\;
h^{++\dot\alpha(s-1)\alpha(s-2)+}(\zeta), 
\end{equation}
          where we used notations for symmetrized combinations of indices $\alpha(s) := (\alpha_1 \ldots \alpha_s), \dot\alpha(s) := (\dot\alpha_1 \ldots \dot\alpha_s)$.
          These superfields are related by rigid $\mathcal{N}=2$ supersymmetry transformations:
          \begin{equation}\label{eq: susy3}
            \begin{split}
          &\delta_\epsilon h^{++\alpha(s-1)\dot\alpha(s-1)} = -4i\big[h^{++\alpha(s-1)(\dot\alpha(s-2)+}\bar\epsilon^{-\dot\alpha_{s-1})}-
          h^{++\dot\alpha(s-1)(\alpha(s-2)+}\,\epsilon^{-\alpha_{s-1})}
          \big]\,,
          \\
          &\delta_\epsilon h^{++\alpha(s-2)\dot\alpha(s-2)} =2i\big[h^{++(\alpha(s-2)\alpha_{s-1})
            \dot\alpha(s-2)+}\epsilon^{-}_{\alpha_{s-1}} +
          h^{++\alpha(s-2)(\dot\alpha(s-2)\dot\alpha_{s-1})+}\,\bar\epsilon^{-}_{\dot{\alpha}_{s-1}}
          \big],
          \\
          &\delta_\epsilon h^{++\alpha(s-1)\dot{\alpha}(s-2)+} = 0,
          \qquad\qquad
              \delta_\epsilon h^{++\dot{\alpha}(s-1)\alpha(s-2)+} = 0.
          \end{split}
          \end{equation}
The gauge potentials are defined up to a gauge freedom:
\begin{equation}\label{eq: spin s gauge transformations}
    \begin{split}
         &\delta_\lambda h^{++\alpha(s-1)\dot\alpha(s-1)} = \mathcal{D}^{++} \lambda^{\alpha(s-1)\dot\alpha(s-1)} +
        4i \big[\lambda^{+\alpha(s-1)(\dot\alpha(s-2)}\bar\theta^{+\dot\alpha_{s-1})}
        \\
        &\qquad\qquad\qquad\qquad\qquad\qquad\qquad\qquad\qquad
        +\,\theta^{+(\alpha_{s-1}} \bar\lambda^{+\alpha(s-2))\dot\alpha(s-1)} \big],
        \\
        &\delta_\lambda h^{++\alpha(s-2)\dot\alpha(s-2)} = \mathcal{D}^{++} \lambda^{\alpha(s-2)\dot\alpha(s-2)} -
        2i\,\big[\lambda^{+(\alpha(s-2)\alpha_{s-1})\dot\alpha(s-2)} \theta^+_{\alpha_{s-1}}
        \\
        &\qquad\qquad\qquad\qquad\qquad\qquad\qquad\qquad\qquad
          +\,
        \bar\lambda^{+(\dot\alpha(s-2)\dot\alpha_{s-1})\alpha(s-2)} \bar\theta^+_{\dot\alpha_{s-1}} \big],
        \\
         &\delta_\lambda  h^{++\alpha(s-1)\dot\alpha(s-2)+} = \mathcal{D}^{++}\lambda^{\alpha(s-1)\dot\alpha(s-2)+},
         \\
         &\delta_\lambda h^{++\dot\alpha(s-1)\alpha(s-2)+} =
         \mathcal{D}^{++}\bar\lambda^{\dot\alpha(s-1)\alpha(s-2)+}.
    \end{split}
\end{equation}

Using this freedom, one can impose the Wess-Zumino gauge:
\begin{equation}\label{eq: WZ gauge}
    \begin{split}
&  h_{WZ}^{++\alpha(s-1)\dot{\alpha}(s-1)}
=
-4i \theta^{+}_{\beta} \bar{\theta}^{+}_{\dot{\beta}} \Phi^{(\beta\alpha(s-1))(\dot{\beta}\dot{\alpha}(s-1))}_{}
-
4i \theta^{+(\alpha} \bar{\theta}^{+(\dot{\alpha}}
\Phi^{\alpha(s-2))\dot{\alpha}(s-2))}
\\
&\qquad\qquad\qquad\qquad
+  (\bar{\theta}^+)^2 \theta^{+\beta} \psi_{\beta}^{\alpha(s-1)\dot{\alpha}(s-1)i}u^-_i
+\, (\theta^+)^2 \bar{\theta}^{+\dot{\beta}} \bar{\psi}_{\dot{\beta}}^{\alpha(s-1)\dot{\alpha}(s-1)i}u_i^-
\\&\qquad\qquad\qquad\qquad
+  (\theta^+)^2 (\bar{\theta}^+)^2 V^{\alpha(s-1)\dot{\alpha}(s-1)(ij)}u^-_iu^-_j\,,
\\
&  h_{WZ}^{++\alpha(s-2)\dot{\alpha}(s-2)} =
-4i \theta^{+}_{\beta} \bar{\theta}^{+}_{\dot{\beta}} C^{(\beta\alpha(s-2))(\dot{\beta}\dot{\alpha}(s-2))}_{}
-4i \theta^{+(\alpha} \bar{\theta}^{+(\dot{\alpha}} C^{\alpha(s-3))\dot{\alpha}(s-3))}
\\&\qquad\qquad\qquad\qquad+ (\bar{\theta}^+)^2 \theta^{+\beta} \rho_\beta^{\alpha(s-2)\dot{\alpha}(s-2)i}u^-_i
+ (\theta^+)^2 \bar{\theta}^{+\dot{\beta}} \bar{\rho}_{\dot{\beta}}^{\alpha(s-2)\dot{\alpha}(s-2)i}u_i^-
\\&\qquad\qquad\qquad\qquad+ (\theta^+)^2 (\bar{\theta}^+)^2 S^{\alpha(s-2)\dot{\alpha}(s-2)(ij)}u^-_iu^-_j\,,
\\
&  h_{WZ}^{++\alpha(s-1)\dot{\alpha}(s-2)+} = (\theta^+)^2 \bar{\theta}^+_{\dot{\beta}} P^{\alpha(s-1)\dot{\alpha}(s-2)\dot{\beta}}
+  \left(\bar{\theta}^+\right)^2 \theta^+_\beta T^{\dot\alpha(s-2)\alpha(s-1)\beta}
\\&\qquad\qquad\qquad\qquad+  (\theta^+)^2 (\bar{\theta}^+)^2 \chi^{\alpha(s-1)\dot{\alpha}(s-2)i}u^-_i\,,
\\
&  h_{WZ}^{++\dot{\alpha}(s-1)\alpha(s-2)+} = \widetilde{\left(h_{WZ}^{++\alpha(s-1)\dot{\alpha}(s-2)+}\right)}\,.
    \end{split}
\end{equation}
These remaining fields form the off-shell $\mathcal{N}=2$ supersymmetric spin $\mathbf{s}$ multiplet. The residual gauge freedom, which preserves the WZ gauge form \eqref{eq: WZ gauge},
implies the appropriate gauge freedom on the component fields. As a result, we obtain a set of physical massless spin fields
\footnote{For a review of description of higher-spin fields in the spinor notation see, e.g.,  \cite{BK}.} $(s, s-1/2, s-1/2, s-1)$ and a set of auxiliary
fields\footnote{Note that some auxiliary fields must be properly redefined in terms of physical ones for ensuring them to be gauge group scalars.}:

$\bullet$ Fields $(\Phi^{\alpha(s)\dot{\alpha}(s)}, \Phi^{\alpha(s-2)\dot{\alpha}(s-2)})$ and $(C^{\alpha(s-1)\dot{\alpha}(s-1)}, C^{\alpha(s-3)\dot{\alpha}(s-3)})$ correspond to the massless Fronsdal spin $s$ and $s-1$ fields:
\begin{equation}
    \begin{split}
    &\delta \Phi^{\alpha(s)\dot{\alpha}(s)} \sim \partial^{(\alpha(\dot{\alpha}} a^{\alpha(s-1))\dot{\alpha}(s-1))},
    \qquad
    \delta \Phi^{\alpha(s-2)\dot{\alpha}(s-2)} \sim \partial_{\beta\dot{\beta}} a^{(\beta\alpha(s-2))(\dot{\beta}\dot{\alpha}(s-2))};
    \\
    &\delta C^{\alpha(s-1)\dot{\alpha}(s-1)} \sim \partial^{(\alpha(\dot{\alpha}} b^{\alpha(s-2))\dot{\alpha}(s-2))},
     \qquad
     \delta C^{\alpha(s-3)\dot{\alpha}(s-3)} \sim \partial_{\beta\dot{\beta}} a^{(\beta\alpha(s-3))(\dot{\beta}\dot{\alpha}(s-3))}.
    \end{split}
\end{equation}

$\bullet$ Fields $V^{\alpha(s-1)\dot{\alpha}(s-1)(ij)}, S^{\alpha(s-2)\dot{\alpha}(s-2)(ij)}$ are real bosonic auxiliary fields,
$P^{\alpha(s-1)\dot{\alpha}(s-2)\dot{\mu}},$ $T^{\alpha(s-1)\nu\dot{\alpha}(s-2)}$ are complex bosonic auxiliary fields.

$\bullet$ Fields $\left( \psi^{\alpha(s-1)\dot{\alpha}(s-1)}_\beta, \bar{\rho}_{\dot{\beta}}^{\alpha(s-2)(\dot{\alpha}(s-3)\dot{\beta})} \right)$ possess gauge freedom
characteristic of the doublet of massless spin $s-\frac{1}{2}$ Fang-Fronsdal fields:
\begin{equation}
    \delta \psi_\beta^{\alpha(s-1)\dot{\alpha}(s-1)} \sim \partial_\beta^{(\dot{\alpha}} \xi^{\alpha(s-1)\dot{\alpha}(s-1))},
    \qquad
    \delta \bar{\rho}^{\alpha(s-2)(\dot{\alpha}(s-3)\dot{\beta})}_{\dot{\beta}}
    \sim
    \partial_{\beta\dot{\beta}} \xi^{(\alpha(s-2)\beta)(\dot{\alpha}(s-3)\dot{\beta})}.
\end{equation}

$\bullet$ Fields $\rho^{\alpha(s-1)\dot{\alpha}(s-2)i}, \chi^{\alpha(s-1)\dot{\alpha}(s-2)i}$ are auxiliary  fermionic fields.

As a result, $\mathcal{N}=2$ spin $\mathbf{s}$ supermultiplet involves $\mathbf{8(s^2+(s-1)^2)}_B+\mathbf{8(s^2+(s-1)^2)}_F$
off-shell degrees of freedom\footnote{On shell there survive $4_B+4_F$ degrees of freedom for each $\mathcal{N}=2$ spin $\mathbf{s}$.}.
In the simplest $s=2$ case one reproduces the off-shell multiplet of the ``minimal''  $\mathcal{N}=2$ Einstein supergravity.

\medskip

The  manifestly $\mathcal{N}=2$ supersymmetric and gauge invariant linearized action has the universal form for any ${\bf s}$:
\begin{equation}\label{eq: spin s action}
    \begin{split}
S_{(s)} = (-1)^{s+1} \int d^4x
d^8\theta du \,\Big\{&G^{++\alpha(s-1)\dot\alpha(s-1)}G^{--}_{\alpha(s-1)\dot\alpha(s-1)}
\\
&\qquad+4 G^{++\alpha(s-2)\dot\alpha(s-2)}G^{--}_{\alpha(s-2)\dot\alpha(s-2)}
\Big\},
\end{split}
\end{equation}
where we have introduced $\mathcal{N}=2$ supersymmetry-covariant fields ($\delta_\epsilon G^{++\dots} = 0$)
\begin{equation}
    \begin{split}
G^{++\alpha(s-1)\dot\alpha(s-1)} = h^{++\alpha(s-1)\dot\alpha(s-1)} + 4i \big[&h^{++\alpha(s-1)(\dot\alpha(s-2)+}\bar\theta^{-\dot\alpha_{s-1})} \\
&- h^{++\dot\alpha(s-1)(\alpha(s-2)+}\,\theta^{-\alpha_{s-1})} \big],  \\
 G^{++\alpha(s-2)\dot\alpha(s-2)} = h^{++\alpha(s-2)\dot\alpha(s-2)} - 2i \big[&h^{++\alpha(s-2)\alpha_{s-1}) \dot\alpha(s-2)+}\theta^{-}_{\alpha_{s-1}}  \\
&+ h^{++\alpha(s-2)(\dot\alpha(s-2)\dot\alpha_{(s-1)})+}\,\bar\theta^{-}_{\dot{\alpha}_{s-1}} \big],
    \end{split}
\end{equation}
and the negatively charged potentials are related to the basic ones by the appropriate harmonic zero-curvature conditions:
\begin{equation}
    \begin{split}
    \mathcal{D}^{++} G^{--\alpha(s-1)\dot{\alpha}(s-1)} = \mathcal{D}^{--} G^{++\alpha(s-1)\dot{\alpha}(s-1)},
    \\
    \mathcal{D}^{++} G^{--\alpha(s-2)\dot{\alpha}(s-2)} = \mathcal{D}^{--} G^{++\alpha(s-2)\dot{\alpha}(s-2)}.
    \end{split}
\end{equation}

Using the explicit form of the WZ gauge \eqref{eq: WZ gauge} and eliminating the auxiliary fields,
one can verify that  the action \eqref{eq: spin s action} at the component level is reduced to the sum of the free Fronsdal actions  for the spins $s$ and $s-1$, as well as  two Fang-Fronsdal actions for the spin $s-1/2$.

\section{$\mathcal{N}=2$ supersymmetric interaction of higher spins with hypermultiplet}
\label{sec: 4}

Using the spin $\mathbf{s}$ analytic gauge potentials \eqref{eq: non-conf prepot}, one can define the analytic differential operator:
\begin{equation}\label{eq: spin s operator}
    \begin{split}
    \hat{\mathcal{H}}^{++}_{(s)}: =& h^{++\alpha(s-1)\dot{\alpha}(s-1)} \partial^{s-1}_{\alpha(s-1)\dot{\alpha}(s-1)}
    +
    h^{++\alpha(s-1)\dot{\alpha}(s-2)+} \partial^{s-2}_{\alpha(s-2)\dot{\alpha}(s-2)}\partial^-_\alpha
    \\&+h^{++\alpha(s-2)\dot{\alpha}(s-1)+}  \partial^{s-2}_{\alpha(s-2)\dot{\alpha}(s-2)}\partial^-_{\dot{\alpha}}
    +
    h^{++\alpha(s-2)\dot{\alpha}(s-2)} \partial^{s-2}_{\alpha(s-2)\dot{\alpha}(s-2)} \partial_5.
    \end{split}
\end{equation}
It is direct to check that this operator is  invariant under $\mathcal{N}=2$ supersymmetry \eqref{eq: susy1}, \eqref{eq: susy2}, \eqref{eq: susy3}. For spin $s$, it contains $s-1$ derivative.

The most general cubic hypermultiplet coupling to spin $\mathbf{s}$ higher-spin consistent with analyticity and  $\mathcal{N}=2$ supersymmetry has the form:
\begin{equation}\label{eq: spin s vertex}
    S_{free} + S_{cubic, s} = - \frac{1}{2} \int d\zeta^{(-4)} \,q^{+a} \left( \mathcal{D}^{++} + \kappa_s\hat{\mathcal{H}}^{++}_{(s)} (J)^{P(s)}\right) q^+_a.
\end{equation}
Here we used the notation $P(s):= \frac{1- (-1)^s}{2}$ and $\kappa_s$ is the spin $\mathbf{s}$ coupling constant. The generator $J$ acts as $J\tilde{q}^+ := i \tilde{q}^+$, $Jq^+ := -i q^+$.
The reason for introducing this generator is that without it
the vertices for odd spins prove to be identically zero\footnote{This is an analogue of the well-known fact that
it is impossible to construct a minimal interaction of spin 1 (and all odd spins) with a real scalar field. Generator $J$ indicates that the fields $q^+$ and $\tilde{q}^+$ ( $f^i$ and $\bar{f}^i$ at the component level)
have the opposite charges.}. So for odd spins this interaction explicitly breaks $SU(2)_{PG}$.

Under the action of gauge transformations \eqref{eq: spin s gauge transformations}, operator $\hat{\mathcal{H}}^{++}_{(s)}$ transforms as:
\begin{equation}
    \delta_\lambda \hat{\mathcal{H}}^{++}_{(s)} = [\mathcal{D}^{++}, \hat{\Lambda}_{(s)}],
\end{equation}
where
\begin{equation}
    \begin{split}
    \hat{\Lambda}_{(s)}:=&
    \lambda^{\alpha(s-1)\dot{\alpha}(s-1)} \partial^{s-1}_{\alpha(s-1)\dot{\alpha}(s-1)}
    +
    \lambda^{\alpha(s-1)\dot{\alpha}(s-2)+}\partial^{s-2}_{\alpha(s-2)\dot{\alpha}(s-2)} \partial^-_\alpha
    \\&+
    \lambda^{\alpha(s-2)\dot{\alpha}(s-1)+} \partial^{s-2}_{\alpha(s-2)\dot{\alpha}(s-2)} \partial^-_{\dot{\alpha}}
    +
    \lambda^{\alpha(s-2)\dot{\alpha}(s-2)} \partial^{s-2}_{\alpha(s-2)\dot{\alpha}(s-2)} \partial_5.
    \end{split}
\end{equation}

One can check that the action \eqref{eq: spin s vertex} is invariant to the leading order in $\kappa_s$, if the gauge transformation of the  hypermultiplet is:
\begin{equation}\label{eq: hyper gauge s}
    \delta^{(s)}_\lambda q^+_a =
    -\frac{\kappa_s}{2} \left\{\hat{\Lambda}_{(s)}^{\alpha(s-2)\dot{\alpha}(s-2)}
    +
    \frac{1}{2}\Omega_{(s)}^{\alpha(s-2)\dot{\alpha}(s-2)}, \partial^{s-2}_{\alpha(s-2)\dot{\alpha}(s-2)} \right\} J^{P(s)} q^+_a.
\end{equation}
Here $\hat{\Lambda}_{(s)}^{\alpha(s-2)\dot{\alpha}(s-2)} = \lambda^{\alpha(s-2)\dot{\alpha}(s-2)M}\partial_M$, $\Omega_{(s)}^{\alpha(s-2)\dot{\alpha}(s-2)} = (-1)^{P(M)} \partial_M\lambda^{\alpha(s-2)\dot{\alpha}(s-2)M}$,
where $M := (\alpha\dot{\alpha}, \alpha+, \dot{\alpha}+)$, and $P(M)$ is defined as $P(\alpha\dot{\alpha})=0, P(\alpha+) = P(\dot{\alpha}+)=1$.

The  cubic vertices constructed are invariant only to the leading order. The variation of the cubic part of the action \eqref{eq: spin s vertex} under \eqref{eq: hyper gauge s}
in the order $\sim \kappa_s^2$ includes terms which are quadratic in spinor derivatives and so cannot be compensated by cubic vertices of such a type.
The only exception is the case of $s=2$, i.e. $\mathcal{N}=2$ Einstein supergravity. In this case, by a non-Abelian deformation of the prepotential gauge transformation,
\begin{equation}
    \delta_\lambda \hat{\mathcal{H}}^{++}_{(s=2)} = [\mathcal{D}^{++}, \hat{\Lambda}_{(s=2)}]
    \quad
    \to
    \quad
    \delta^{full}_\lambda \hat{\mathcal{H}}^{++}_{(s=2)} = [\mathcal{D}^{++}+ \kappa_2 \hat{\mathcal{H}}^{++}_{(s=2)}, \hat{\Lambda}_{(s=2)}],
\end{equation}
the complete invariance of the action \eqref{eq: spin s vertex} can be ensured.

Perhaps for $s\geq3$ a non-linear gauge invariance can be achieved by introducing a new type of cubic vertices and a new type of gauge superfield potentials.
However, the meaning of such potentials, their gauge freedom and their field contents are unclear at present and require further study.
It is not unlikely that they are nontrivially
related to $\mathcal{N}=2$ gauge potentials of the half-integer higher spins, which still remain to be constructed. By introducing an infinite tower of $\mathcal{N}=2$ higher-spin prepotentials
(with both integer and half-integers spins), one could expect the complete non-Abelian invariance.
It is interesting to figure out possible parallels of these assumptions with the relevant discussions in \cite{Bekaert:2009ud}.

Note that, for a special choice of the transformation parameters\footnote{Such a set of parameters is a $\mathcal{N}=2$ supersymmetric generalization of the Killing tensor field
(see, e.g., \cite{Bekaert:2009ud, Howe:2015bdd}) and can be interpreted as a ``$\mathcal{N}=2$ Killing supertensor''.}, such that $[\mathcal{D}^{++}, \hat{\Lambda}_{(s)}] = 0$,
we obtain rigid ``higher-spin''
supersymmetry transformations of the free hypermultiplet action with $s-1$ derivatives\footnote{In the $s=2$ case these transformations correspond to rigid $\mathcal{N}=2$ supersymmetry.}.
Thus the  interactions constructed have a Noether origin and so can be viewed as gauging of the ``higher-spin'' rigid supersymmetries of the free hypermultiplet
through introducing the appropriate analytic gauge potentials. Also it is possible to reconstruct both gauge transformations and supersymmetry transformations
of the $\mathcal{N}=2$ higher-spin potentials.
All these reasonings will be helpful while constructing $\mathcal{N}=2$ superconformal higher-spin gauge potentials in the next section.

\section{$\mathcal{N}=2$ superconformal higher-spin multiplets}
\label{sec: 5}

$\mathcal{N}=2$ superconformal transformations non-trivially act on harmonics \cite{18} (in contrast to $\mathcal{N}=2$ supersymmetry, see \eqref{eq: susy1}). So we extend the set
of indices $M$ by harmonic indices $\Rightarrow \; M :=(\alpha\dot{\alpha}, \alpha+, \dot{\alpha}+, ++)$, $P(++)=0$.
Free off-shell action of massless hypermultiplet \eqref{eq: hyper free}  is invariant under $\mathcal{N}=2$ superconformal transformations (here it is useful to employ a passive form of transformations):
\begin{equation}\label{eq: sc hyp}
    \delta_{sc} q^+_a = - \hat{\Lambda} q^+_a - \frac{1}{2} \Omega q^+_a,
    \qquad
    \hat{\Lambda} = \lambda^M\partial_M,
    \quad
    \Omega = (-1)^M \partial_M \lambda^M
\end{equation}
with special rigid parameters:
\begin{equation}\label{eq:superconformal symmetry}
    \begin{cases}
        \lambda_{sc}^{\alpha\dot{\alpha}}
        =&
        a^{\alpha\dot{\alpha}}
        -
        4i \left( \epsilon^{\alpha i} \bar{\theta}^{+\dot{\alpha}} + \theta^{+\alpha} \bar{\epsilon}^{\dot{\alpha}i} \right) u^-_i

        +
        x^{\dot{\alpha}\rho} k_{\rho\dot{\rho}} x^{\dot{\rho}\alpha}
        + b x^{\alpha\dot{\alpha}}
        \\&-
        4i \theta^{+\alpha} \bar{\theta}^{+\dot{\alpha}} \lambda^{(ij)}u^-_i u^-_j
        -
        4i \left(x^{\alpha\dot{\rho}}\eta_{\dot{\rho}}^i \bar{\theta}^{+\dot{\alpha}}
        +
        \theta^{+\alpha} \eta^i_{\rho} x^{\rho\dot{\alpha}}
        \right) u^-_i,
        \\
        \lambda^{\alpha+}_{sc}
        =&
        \epsilon^{\alpha i} u^+_i
        +
        \frac{1}{2}
        \theta^{+\alpha} (b + i \gamma)
        +
        x^{\alpha\dot{\beta}} k_{\beta\dot{\beta}} \theta^{+\beta}
        +
        x^{\alpha\dot{\alpha}}  \eta^i_{\dot{\alpha}} u^+_i
        \\&
        +
        \theta^{+\alpha}
        \left( \lambda^{(ij)}u^+_i u^-_j
        +
        4i \theta^{+\rho} \eta^i_{\rho} u^-_i
        \right),
        \\
        \bar{\lambda}^{\dot{\alpha}+}_{sc}
        =&
        \epsilon^{\dot{\alpha} i} u^+_i
        +
        \frac{1}{2}
        \bar{\theta}^{+\dot{\alpha}}  (b - i \gamma)
        +
        x^{\dot{\alpha}\beta} k_{\beta\dot{\beta}} \bar{\theta}^{+\dot{\beta}}
        +
        x^{\alpha\dot{\alpha}}  \eta^i_{\alpha} u^+_i
        \\&
        +
        \bar{\theta}^{+\dot{\alpha}}
        \left( \lambda^{(ij)}u^+_i u^-_j
        -
        4i \bar{\theta}^{+\dot{\rho}} \eta^i_{\dot{\rho}} u^-_i
        \right),
        \\
        \lambda^{++}_{sc} =&
        \lambda^{ij} u^+_i u^+_j
        +
        4i \theta^{+\alpha} \bar{\theta}^{+\dot{\alpha}} k_{\alpha\dot{\alpha}}
        +
        4i \left( \theta^{+\alpha} \eta^i_{\alpha} + \eta^i_{\dot{\alpha}} \bar{\theta}^{+\dot{\alpha}}  \right) u^+_i.
    \end{cases}
\end{equation}
These parameters satisfy the system of equations $[\mathcal{D}^{++}, \hat{\Lambda}] = \lambda^{++}\mathcal{D}^0$.

Following the discussion in the end of previous section, one can determine $\mathcal{N}=2$ higher-spin superconformal gauge potentials
by requiring superconformal invariance of general cubic vertex\footnote{An alternative approach to determining the structure
of $\mathcal{N}$-extended superconformal higher-spin gauge prepotentials (of Mezincescu type) uses the method of supercurrent
multiplets and starts from the Fayet-Sohnius on shell hypermultiplets \cite{Kuzenko:2021pqm}. Our approach,
like in the non-conformal case \cite{Buchbinder:2021ite} - \cite{Buchbinder:2022vra},  uses the analytic gauge vielbein-type potentials covariantizing the ${\cal D}^{++}$ derivative,
and the off-shell $q^+$ hypermultiplets as
most general ${\cal N}=2$ matter multiplets. An interesting task is to learn how $\mathcal{N}=2$ Kuzenko-Raptakis prepotentials emerge from our analytic potentials
in a special harmonic-independent gauge.}.
The analyticity and superconformal invariance taken together imply the following  general form of cubic interaction:
\begin{equation}\label{eq: spin s sc vertex}
    S_{free} + S_{sc-cubic,s}
    =
    -
    \frac{1}{2} \int d\zeta^{(-4)}\,
    q^{+a} \left(\mathcal{D}^{++} + \hat{\mathbb{H}}^{++}_{(s)} (J)^{P(s)} \right) q^+_a,
\end{equation}
where we introduced the analytic differential operator of degree  $s-1$ with the odd number of superspace derivatives
\begin{equation}\label{eq: sc spin s operator}
    \hat{\mathbb{H}}^{++}_{(s)} := h^{++M_1\dots M_{s-1}} \partial_{M_{s-1}} \dots \partial_{M_1}
    +
    h^{++M_1\dots M_{s-3}} \partial_{M_{s-3}} \dots \partial_{M_1}
    +
    \dots.
\end{equation}
Various parts of the operator can be viewed as corresponding to the spins $\mathbf{s}, \mathbf{s-2}, \dots$.
This operator generalizes the non-conformal operator $\hat{\mathcal{H}}^{++}_{(s)}$ given in \eqref{eq: spin s operator} in three aspects.
First of all, it contains all possible types of indices since the derivative $\partial_{\alpha\dot{\alpha}}$  is non-trivially transformed under the superconformal group \eqref{eq:superconformal symmetry}.
Secondly, because of the presence of a weight factor in \eqref{eq: sc hyp}, terms with fewer derivatives are necessarily present. Thirdly, no the irreducibility conditions with respect to the Lorentz indices
are imposed in advance on the potentials, as distinct from a symmetrization in the non-conformal case \eqref{eq: non-conf prepot}.

From now on, to avoid identical terms, we assume that the indices of the fields are ordered according to the rule $M =(\alpha\dot{\alpha}, \alpha+, \dot{\alpha}+, ++)$, i.e.
$M_1\geq M_2\dots \geq M_{s-1}$ and prepotentials satisfy the permutation condition $h^{++M_1\dots M_n M_k\dots M_r} = (-1)^{P(M_k)P(M_n)} h^{++M_1\dots M_kM_n \dots M_r} $.

For lucidity, we give the explicit form of the superconformal transformation law of  higher-spin prepotentials, which ensures the superconformal invariance of the cubic vertex:
\begin{equation}\label{GenScGauge}
    \delta_{sc}     \hat{\mathbb{H}}^{++}_{(s)}
    =
    [   \hat{\mathbb{H}}^{++}_{(s)}, \hat{\Lambda}]
    +
    \frac{1}{2} [   \hat{\mathbb{H}}^{++}_{(s)}, \Omega].
\end{equation}
Here we assume the appropriate integration by parts which brings the right side of \eqref{GenScGauge} to the form \eqref{eq: sc spin s operator}.
All the peculiarities listed above are in conjunction with this formula.

The gauge freedom of action \eqref{eq: spin s sc vertex} is realized by the transformations:
\begin{subequations}\label{eq: gauge conf}
\begin{equation}
    \delta^{(s,k)}_\lambda q^+_a = -
    \frac{\kappa_s}{2} \left\{ \hat{\Lambda}^{M_1\dots M_{k-2}} + \frac{1}{2}\Omega^{M_1\dots M_{k-2}}, \partial_{M_{k-2}}\dots \partial_{M_1} \right\}_{AGB} (J)^{P(s)} q^+_a,
\end{equation}
\begin{equation}\label{eq: sc spin s transformations}
    \delta^{(s,k)}_\lambda  \hat{\mathbb{H}}^{++}_{(s)} =
    \frac{1}{2} \left[\mathcal{D}^{++}, \left\{ \hat{\Lambda}^{M_1\dots M_{k-2}}, \partial_{M_{k-2}}\dots \partial_{M_1} \right\}_{AGB} \right],
\end{equation}
\end{subequations}
where in \eqref{eq: sc spin s transformations} we performed the appropriate integration by parts. We also define operators $\hat{\Lambda}^{M_1\dots M_{k-2}} : = \lambda^{M_1\dots M_{k-2}N}\partial_N$,
$\Omega^{M_1\dots M_{k-2}} : = (-1)^{P(N)}\partial_N\lambda^{NM_1\dots M_{k-2}}$. The $k$-th parameter takes the values $s,s-2.\dots$. The different values of $k$ correspond to the gauge freedom
for different spin contributions appearing in the operator \eqref{eq: sc spin s operator}. Anti-graduate bracket is defined as $\{F_1, F_2\}_{AGB} := [F_1, F_2]$ for fermionic objects, $\{B_1, B_2\}_{AGB} := \{B_1, B_2\}$ for bosonic and $\{B, F\}_{AGB} := \{B, F\}$ for bosonic and fermionic ones.

If we require $q^{+a}   \delta^{(s,k)}_\lambda  \hat{\mathbb{H}}^{++}_{(s)}  (J)^{P(s)}q^+_a =0$, we obtain the conditions on ``superconformal higher-spin'' rigid symmetries
of the free hypermultiplet (in the $s=2$ case these symmetry transformations are just $\mathcal{N}=2$ superconformal transformations). The corresponding parameters $\lambda^{\dots}$
which solve this equation can be treated as $\mathcal{N}=2$ spin $s$ ``superconformal Killing supertensor''.

Using the gauge freedom \eqref{eq: sc spin s transformations}, one can impose WZ gauge\footnote{For the $s=3$ case we have derived this result directly, by exploring
the gauge freedom \eqref{eq: sc spin s transformations}. The WZ gauge for general $s$ was conjectured as a natural generalization of the spin $3$ WZ gauge.}:
\begin{equation}
        \begin{split}
            &h_{WZ}^{++\alpha(s-1)\dot{\alpha}(s-1)}
            =
            -4i \theta^{+}_{\rho}\bar{\theta}^{+}_{\dot{\rho}} \Phi^{(\rho\alpha(s-1))(\dot{\rho}\dot{\alpha}(s-1))}
            -
            (\bar{\theta}^+)^2 \theta^{+}_{\rho} \psi_{}^{(\rho\alpha(s-1))\dot{\alpha}(s-1)i} u_i^-
            \\&\;\;\;\;\qquad\qquad\qquad\quad-
            (\theta^+)^2 \bar{\theta}^{+}_{\dot{\rho}} \bar{\psi}^{\alpha(s-1)(\dot{\alpha}(s-1)\dot{\rho})i} u_i^-
            +
            (\theta^+)^2 (\bar{\theta}^+)^2 V^{\alpha(s-1)\dot{\alpha}(s-1)ij}u^-_i u^-_j\,,\\
            &h_{WZ}^{++\alpha(s-1)\dot{\alpha}(s-2)+}
            =
            (\theta^{+})^2 \bar{\theta}^{+}_{\dot{\nu}} P^{\alpha(s-1)(\dot{\alpha}(s-2)\dot{\nu})}
            +
            (\bar{\theta}^{+})^2 \theta^{+}_{\nu} T^{(\alpha(s-1)\nu)\dot{\alpha}(s-2)}_{}
            \\&\qquad\qquad\qquad\qquad\qquad\quad\qquad\qquad\qquad\quad\;\;\,+
            (\theta^{+})^4 \chi^{\alpha(s-1)\dot{\alpha}(s-2)i}u_i^-\,,
            \\
            &h_{WZ}^{++\alpha(s-2)\dot{\alpha}(s-1)+}=
            \widetilde{h_{WZ}^{++\alpha(s-1)\dot{\alpha}(s-2)+}}\,,
            \\&h_{WZ}^{(+4)\alpha(s-2)\dot{\alpha}(s-2)} =
            (\theta^+)^2 (\bar{\theta}^+)^2 D^{\alpha(s-2)\dot{\alpha}(s-2)}\,.
        \end{split}
\end{equation}
From \eqref{eq: sc spin s transformations} it also follows, that all other potentials are purely gauge degrees of freedom. In the gauge that we have fixed,
they are set to zero. In principle, it is possible to initially fix the gauge in which these fields
are zero, but in this case the superconformal group will be implemented non-linearly.
This situation is typical for the conformal theory of higher spins, see, e.g., \cite{Kuzenko:2022hdv}.

The $\mathcal{N}=2$ superconformal higher-spin  multiplet displays a number of new features in comparison to the non-conformal case.
First of all, there are no auxiliary fields in the multiplet for $s\geq 3$ -- all fields prove to be the gauge fields. Secondly, the physical degrees of freedom for each spin in the multiplet
are described by one irreducible field and their gauge transformation laws have a slightly different form\footnote{Some fields require proper redefinitions to avoid contributions of other fields
in the relevant transformation laws}:

$\bullet$ Bosonic fields $\Phi^{\alpha(s)\dot{\alpha}(s)}$,
$V^{\alpha(s-1)\dot{\alpha}(s-1)ij}$, $P^{(\alpha(s-1)\dot{\alpha}(s-1))}$,  $D^{\alpha(s-2)\dot{\alpha}(s-2)}$ are conformal Fradkin-Tseytlin fields \cite{Fradkin:1985am} with gauge freedom of the form:
\begin{equation}
    \delta \Phi^{\alpha(s)\dot{\alpha}(s)}
    \sim
    \partial^{(\alpha(\dot{\alpha}} a^{\alpha(s-1))\dot{\alpha}(s-1))},
    \quad
    \delta V^{\alpha(s-1)\dot{\alpha}(s-1)ij}
    \sim \partial^{(\alpha(\dot{\alpha}} v^{\alpha(s-2))\dot{\alpha}(s-2))(ij)}, \dots.
\end{equation}

$\bullet$ Complex field $T^{\alpha(s)\dot{\alpha}(s-2)}$ have the gauge freedom:
\begin{equation}
    \delta T^{\alpha(s)\dot{\alpha}(s-2)} \sim \partial^{(\alpha(\dot{\alpha}} t^{\alpha(s-1))\dot{\alpha}(s-3))}.
\end{equation}
This is more general type of conformal field, see, e.g., \cite{Kuzenko:2017ujh}. In the simplest non-trivial $s=3$ case this field is equivalent to
the traceless hook gauge field $T^{a[bc]}$ with the algebraic symmetries corresponding to the simplest hook Young diagram $\tiny\yng(2,1)$, see, e.g., \cite{Kuzenko:2020jie}.

$\bullet$ Fields $\psi^{\alpha(s)\dot{\alpha}(s-1)i}$ and $\chi^{\alpha(s-1)\dot{\alpha}(s-2)i}$ are conformal fermionic gauge fields:
\begin{equation}
    \delta \psi^{\alpha(s)\dot{\alpha}(s-1)i}
        \sim \partial^{(\alpha(\dot{\alpha}} \xi^{\alpha(s-1))\dot{\alpha}(s-2))i},
        \quad
        \delta \chi^{\alpha(s-1)\dot{\alpha}(s-2)i}
        \sim
        \partial^{\alpha\dot{\alpha}} \zeta^{\alpha(s-2))\dot{\alpha}(s-3))i}.
\end{equation}
Finally, the $\mathcal{N}=2$ superconformal spin $\mathbf{s}$ multiplet is encompassed by  $\mathbf{8(2s-1)_B+8(2s-1)_F}$ off-shell degrees of freedom.
The simplest $s=1,2$ multiplets (in contrast to $s\geq3$) contain auxiliary fields and comprise $\mathcal{N}=2$ Maxwell and conformal supergravity (Weyl) gauge multiplets.

The action \eqref{eq: spin s sc vertex} with gauge transformations \eqref{eq: gauge conf} is consistent in the first order in $\kappa_s$.
In the conformal case one can construct a fully consistent coupling of an infinite tower of higher-spin potentials to the hypermultiplet by applying to
a non-abelian deformation of gauge freedom.  Also, the action constructed in this way can be consistently  lifted to an  arbitrary background of $\mathcal{N}=2$ superconformal gravity.
More detailed discussion of these issues will be presented in \cite{2024}.

\section{Conclusion}
\label{sec: 6}

In conclusion, we have found that the principle of Grassmann harmonic analyticity plays a key role both in the off-shell formulation of $\mathcal{N}=2$ non-conformal multiplets
of higher spins and $\mathcal{N}=2$ superconformal higher-spins. All the fundamental gauge potentials are analytic superfields, which, being combined with $\mathcal{N}=2$ supersymmetry and conformal supersymmetry ,
significantly limits the possible form of the interaction Lagrangians, as the example of a cubic vertex with a hypermultiplet demonstrates.
This opens up a wide range of new tasks. Let us mention some:

\begin{itemize}
    \item \textit{Massive $\mathcal{N}=2$ higher spins}

    Surprisingly, the problem of constructing supersymmetric massive theories of higher spins is more difficult compared to massless theories \cite{Zinoviev:2007js}.
    The off-shell $4D, \mathcal{N}=1$ superfield theory of massive higher spins has been elaborated quite recently \cite{Koutrolikos:2020tel}. The formulations
    of such theories look rather cumbersome. One would expect that $\mathcal{N}=2$ theory of massive higher spins in HSS would have more elegant formulation.

    \item \textit{Induced actions}

    The vertices \eqref{eq: spin s vertex} and \eqref{eq: spin s sc vertex}  can be used to construct effective theories after calculation
    of the functional integral over hypermultiplets (see, e.g., \cite{Bekaert:2010ky} for a general discussion of higher-spin effective actions).
    This would provide a way of building consistent theories of (superconformal) $\mathcal{N}=2$ higher spins.

    \item \textit{Component reduction}

    The non-conformal vertex \eqref{eq: spin s vertex} is completely self-consistent for $s=2$. The infinite tower
    of $\mathcal{N}=2$ superconformal higher spins interacting with the hypermultiplet is also completely self-consistent, as discussed above.
    The constructed vertices are cubic in superfields. An interesting task is the component reduction of such theories and exploration of the mechanism
    of eliminating the hypermultiplet auxiliary fields, leading to nonlinearities at the component level (see recent \cite{IvAS} for the related discussion).

\end{itemize}

\textbf{Acknowledgements}

The authors thank S.M.~Kuzenko  and E.S.N.~Raptakis for useful correspondence.
Work of N.Z. was partially supported by the grant 22-1-1-42-2 from the Foundation for the Advancement of Theoretical Physics and Mathematics ``BASIS''.

\end{document}